\documentstyle[12pt]{article}

\input{epsf.sty}
\def\framegraphics{\def\ifframe{\iftrue}}
\def\dontframegraphics{\def\ifframe{\iffalse}}
\def\drawgraphics{\def\ifdraw{\iftrue}}
\def\dontdrawgraphics{\def\ifdraw{\iffalse}}

\dontframegraphics
\drawgraphics

\newcommand{\graphics}[6]{
\def\epsfsize##1##2{#6##1}
\begin{picture}(#2,#3)
  \ifframe
    \put(0,0){\framebox(#2,#3){}}
  \fi
  \ifdraw
    \put(0,#3){\begin{picture}(0,0)
                 \put(#4,#5){\epsfbox{#1}}
               \end{picture}}
  \fi
\end{picture}}

\newcommand{\be}{\begin{equation}}
\newcommand{\ee}{\end{equation}}
\newcommand{\bear}{\begin{eqnarray}}
\newcommand{\ear}{\end{eqnarray}}

\begin{document}

ADP-96-12/T215  \hspace*{6cm} February 1996

\vspace{2cm}
\centerline{\large On the Interpretation of the
NA51 Experiment}
\vspace{.4cm}
\centerline{F.M. Steffens and A.W. Thomas}
\vspace{1cm}
\centerline{Department of Physics and Mathematical Physics}
\centerline{University of Adelaide}
\centerline{Adelaide, S.A. 5005, Australia}
\vspace{.3cm}
\begin{abstract}
We study the $p-n$ Drell-Yan asymmetry, recently measured by the NA51
collaboration,
and conclude that the value quoted by their experiment only sets 
a lower limit on the asymmetry of the proton sea.
In particular, we notice that charge symmetry breaking between the 
proton and the neutron may produce corrections
which should be taken into account.

\end{abstract}

The confirmation by the New Muon Collaboration (NMC) \cite{nmc}
of an earlier hint in SLAC data \cite{slac} of a sizeable 
violation of the Gottfried sum-rule \cite{gottfried} has
stimulated enormous interest in the flavour structure of the 
parton distributions \cite{achar}. Early theoretical work by 
Feynman and Field \cite{feynman} had suggested that the 
Pauli Exclusion Principle may lead to $\overline d > \overline u$,
a result supported by calculations using the MIT bag model \cite{signal89}.
However, most of the work aimed at interpreting the NMC results
has been based on the role the pion cloud of the nucleon, which
as first noted in 1983, predicts $\overline d > \overline u$ \cite{tony83}.

Of course, the interpretation of the NMC results in terms of a
violation of flavour symmetry in the sea is not unambiguous. One 
must correct for shadowing and meson cloud contributions 
in the deuteron in order to extract the neutron structure 
function. Nevertheless, there seems to be a consensus that these corrections
enhance the violation of the GSR a little \cite{achar3}. It has also 
been suggested that the violation may be merely apparent, because the 
expected Regge behaviour at small-$x$ may set in less rapidly 
than is usually assumed \cite{mrs90}. In any case, it is very important
to look for other means to test whether in fact $\overline d > \overline u$.

Some time ago, the NA51 group \cite{na51} released a measurement
of the $p - n$ cross section asymmetry defined as

\be
A_{DY} (x) = \frac{\sigma^{pp}(x) - \sigma^{pn}(x)}
{\sigma^{pp}(x) + \sigma^{pn}(x)}.
\ee
This experiment followed a suggestion by Ellis and Stirling
\cite{ellis} where it was argued that the sign and the size of $A_{DY}$ 
can tell whether the sea is symmetric in flavour or not. 
This would be possible because 
$\sigma^{pN} (x) \propto\sum_i e_i^2 (q_i^p (x)\overline{q}_i^N (x) +
\overline{q}_i^p (x)q_i^N (x))$ and then $A_{DY}$ can be expressed as:

\bear
A_{DY} (x) &=& \left\{(4\lambda_v (x) - 1)(\lambda_s (x) - 1) + (\lambda_v (x) - 1)
(4\lambda_s (x) - 1) \right. \nonumber \\*
 && \left. + 2\overline d (x)(4\lambda_s (x) - 1)(\lambda_s (x) - 1)/d_v (x)\right\}/
\nonumber \\*
&& \left\{(4\lambda_v (x) + 1)(\lambda_s (x) + 1) + (\lambda_v (x) + 1)(4\lambda_s (x) + 1)
\right. \nonumber \\*
&& \left. + 2\overline d (x)(4\lambda_s (x) + 1)(\lambda_s (x) + 1)/d_v (x)\right\},
\label{0000}
\ear
with $\lambda_v (x) = u_v(x) /d_v (x)$ and $\lambda_s (x) =
\overline u (x) / \overline d(x)$. For completeness, we have included 
sea-sea corrections in Eq. (\ref{0000}). It is then clear that for a
sea which is SU(2) flavour symmetric, i.e. 
$\lambda_s = 1$, the asymmetry is always positive if $\lambda_v$
is larger than unity. On the other hand the asymmetry can change sign
for $\lambda_s \neq 1$. In particular, if $\lambda_v =2$ the 
asymmetry is negative for $\lambda_s < 0.72$ - where, for simplicity,
the last term (the sea-sea term) was neglected. However, the important 
feature of the $p-n$ cross section asymmetry is that, given 
a value for $\lambda_v$, the measured asymmetry determines whether or not there
is any sort of isospin breaking. In this letter we will explore the
idea that the measurement of $A_{DY}$ is not enough to precisely determine
$\lambda_s$ and thus the degree of asymmetry in the quark sea,
as has been claimed \cite{na51,ellis}.
This point will be clear from  
expression (\ref{05}), where possible corrections from charge symmetry breaking
between the neutron and the proton at a given $x$ are included.

The NA51 collaboration quoted the following result \cite{na51}:

\be
A_{DY} (x = 0.18) = -0.09\pm 0.02(stat) \pm 0.025 (syst)
\label{e1}
\ee
from which they derived

\be
\lambda_s (x=0.18) = 0.51 \pm 0.04 (stat) \pm 0.05 (syst),
\label{01}
\ee
where sea-sea corrections were included but nuclear effects 
were left out. In this interpretation, the experiment indicates 
that there is a strongly asymmetric sea 
at $x=0.18$. We are now going to show that, in fact, the
$\lambda_s$ quoted in Eq. (\ref{01}) is only a limiting value set
within the framework of Eq. (\ref{0000}) which was based on the
assumption of charge symmetry (e.g.
$\overline{u}^p (x) = \overline{d}^n (x)$ and $\overline{d}^p (x) = 
\overline{u}^n (x)$). We could as well have set the 
sea to be flavour symmetric and derived the following expression
for the Drell-Yan asymmetry:

\begin{eqnarray}
A_{DY}(x)&=&\left\{3(\lambda_v (x) - 1) -  (1 + 4\lambda_v (x))\overline\delta (x)/
\overline q (x) + 3 \delta (x)/d_v (x) \right. \nonumber \\*
&& \left. - 10 \overline\delta (x)/d_v (x) \right\}/ \nonumber \\*
&&\left\{13\lambda_v (x) + 7 + (4\lambda_v (x) + 1)\overline\delta (x)/\overline q (x)
- 3 \delta (x)/d_v (x) + \right. \nonumber \\*
&&\left. 10\overline\delta (x)/d_v (x) + 20\overline q (x)/d_v (x)  \right\}.
\label{02}
\end{eqnarray}
Here, we used $\overline u(x) = \overline{u}^p (x) =
\overline{d}^n (x) - \overline\delta (x)$,
$\overline d (x) = \overline{d}^p (x)= \overline{u}^n - 
\overline\delta (x)$, $u_v (x) = u_v^p (x) = d_v^n (x) - \delta (x)$
and $d_v (x) = d_v^p (x) = u_v^n (x) + \delta (x)$. The function $\delta(x)$
does not need to be the same for $u_v$ and $d_v$ but we use the same
function to simplify the expressions as we illustrate the main idea. 
Also notice that different signs are used for the corrections in
$d_v^n$ and $u_v^n$ as this seems to be suggested by theoretical
evidence \cite{rodionov}. We have also considered the
case where both signs coincide and the conclusions presented here 
are insensitive to such a choice.
Of course, $\int_0^1 dx \delta (x) = 0$
to preserve the number of valence quarks. 
Using the experimental result for $A_{DY} (x)$, we can 
estimate the amount of charge symmetry breaking. 

For that purpose, we 
will work with the MRS parametrization, $S_0^{\prime}$, for $\overline q (x)$,
$u_v (x)$ and $d_v (x)$. In this parametrization the sea is symmetric and,
for $x=0.18$, $Q^2 = (5.22\;GeV)^2 = x^2 s$ for $s=(29\; GeV)^2$ the square of 
the center of 
mass energy of the NA51 experiment, it gives $\overline q (x=0.18) = 0.348$, 
$u_v (x=0.18)=3.13$ and $d_v (x=0.18)=1.486$. From the experimental result 
quoted in Eq. (\ref{e1}) we then obtain:
 
\be
\overline\delta (x=0.18) = 0.2088 - 0.0933 \delta (x=0.18).
\label{03}
\ee
In calculating Eq. (\ref{03}) we 
disregarded the sea-sea correction term ($20\overline q (x)/d_v (x)$)
and took only the central
value of the measured asymmetry. This is a good approximation as,
using the same procedure when recalculating result (\ref{01}), we get
$\lambda_s (x=0.18) \simeq 0.53$. Eq. (\ref{03}) tells us that
the interpretation of the Drell-Yan asymmetry purely in terms of
charge symmetry violation is very unlikely because of the size of the
breaking necessary to fit the data. Of course, the procedure is not
entirely consistent because the $S_0^{\prime}$ parametrization was constructed with 
the assumption $\overline\delta (x) = \delta (x) = 0$. Thus, Eq. (\ref{03})
should be seen only as a guide. 

If we take for 
instance $\delta (x) = \overline\delta (x)$, we have 
$\overline\delta (x=0.18)\simeq 0.19$,
which means that the factor giving the breaking is about $55\%$
of the antiquark distribution itself - clearly too large value. 
On the other hand,
there is no reason at all to interpret the NA51 result solely in terms
of isospin breaking between the proton and the neutron.
In the general case, the Drell-Yan asymmetry would be:

\begin{eqnarray}
A_{DY} (x) &=& 
 \left\{(4\lambda_v (x) - 1 )(\lambda_s (x) 
- 1) + (4\lambda_s (x) - 1)(\lambda_v (x) - 1) \right. \nonumber \\*
&& \left.- (4\lambda_v (x) + 1)\overline\delta (x)/\overline d (x) - 
(1 - 4\lambda_s (x))\delta (x)/d_v (x)\right. \nonumber \\*
&& \left. - 2 (4\lambda_s (x) + 1) - 2(4\lambda_s (x) - 1)
(\lambda_s (x) - 1)\overline d (x)/d_v(x)\right\}/ \nonumber \\*
&& \left\{(4\lambda_v (x) + 1 )(\lambda_s (x) 
+ 1) + (4\lambda_s (x) + 1)(\lambda_v (x) + 1) \right. \nonumber \\*
&& \left. + (4\lambda_v (x) + 1)\overline\delta (x)/\overline d (x) + 
(1 - 4\lambda_s (x))\delta (x)/d_v (x)\right. \nonumber \\*
&&\left. + 2 (4\lambda_s (x) + 1) +
2(4\lambda_s (x) + 1)(\lambda_s (x) + 1)\overline d (x)/d_v(x) \right\}.
\label{05}
\end{eqnarray}
To write Eq. (\ref{05}) we made the simplification that,
even for broken sea flavour symmetry, the correction $\overline\delta (x)$
from charge symmetry breaking in the sea has the same form for the $\overline u$ and
for the $\overline d$ quarks. Of course, this does not necessarily need to be the
case.

\begin{figure}[htb]
\graphics{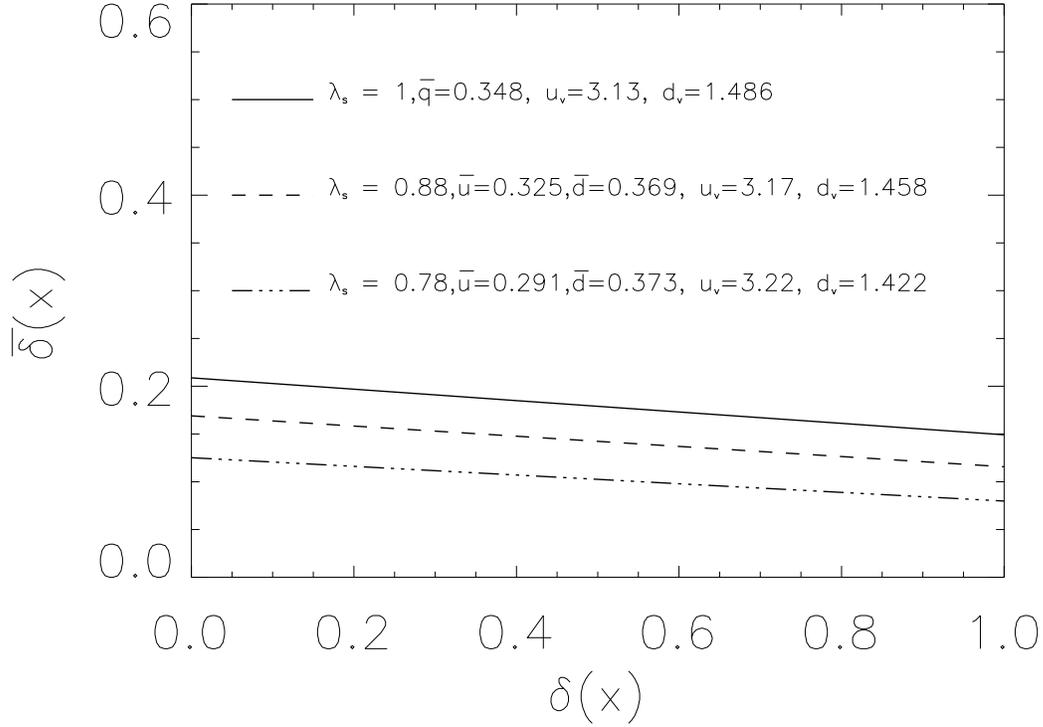}{20}{10}{-2}{-10}{0.9}
\caption{$\overline\delta (x)$ as a  function of $\delta (x)$ for the various
parametrizations discussed in the text.}
\label{figb1}
\end{figure}

Again, to extract any number from Eq. (\ref{05}) we need 
to know the value
of the sea and valence quark distributions at a given $x$.
As we include isospin breaking terms, we have the problem
that there is no standard quark distribution including these corrections.
Moreover, the term involving $\overline\delta (x)$ is potentially 
important as it is multiplied
by a large factor and divided by a small number (viz. $\overline d$). 
This is true whether the integral over $x$ of $\overline\delta (x)$ 
(and $\delta (x)$) is zero or not. This means
that to extract $\lambda_s (x)$ using the measured Drell-Yan asymmetry is
at best ambiguous. To estimate the order of magnitude of $\delta (x)$
and of $\overline\delta (x)$, 
we will assume that the quark and antiquark
distributions are described by the $D_0^{\prime}$ and $D_-^{\prime}$ 
parametrizations.
For the $D_0^{\prime}$ set, the Gottfried sum rule is
0.26 and for the $D_-^{\prime}$ this value is 0.24, which means that
for both sets, $\int_0^1 \overline\delta (x) dx \simeq 0$, but this does 
not mean that $\overline\delta (x)=0$ at 
$x=0.18$. Using the measured asymmetry and disregarding 
sea-sea corrections, one gets:

\begin{eqnarray}
\overline\delta (x=0.18) &=& 0.169 - 0.053 \delta (x=0.18),\;\;
\lambda_s \sim 0.88, \;\;  D_0^{\prime}
\nonumber \\*
\overline\delta (x=0.18) &=& 0.125 - 0.045 \delta (x=0.18).\;\;
\lambda_s \sim 0.78, \;\;  D_-^{\prime}
\label{06}
\end{eqnarray} 
In fig. \ref{figb1} we show the behaviour of $\overline\delta (x)$ 
as a function of $\delta (x)$ for the various parametrizations discussed.
A few comments are in place. First, we see that $\overline\delta$ is not strongly
dependent on $\delta$ and this dependence becomes weaker as $\lambda_s$ decreases.
Moreover, we see that for $\lambda_s=0.78$,
the charge symmetry breaking is of the order $30\% \sim 40\%$ of the
antiquark distribution. Although this value is high, it is at a specific 
value of $x$
and we remember that $\lambda_s=0.78$ is a correction of about $50\%$ to
the central value of $\lambda_s=0.51$ quoted by the NA51 group. We could be
less drastic and propose corrections of the order of $20\%$, bringing the
measured central value to $\lambda_s=0.6$, which is itself a huge correction,
providing a sensitive test for 
any model trying to describe the flavour of the nucleon sea.
As a consequence, if we make a linear extrapolation of $\overline\delta (x)$ with
$\lambda_s$ as suggested by Fig. \ref{figb1}, this correction of $20\%$ would correspond 
to a value of $\overline\delta$ around $10\%$ of the antiquark distribution itself. This could well 
be possible and, from this point of view, the $\lambda_s$ quoted as being experimentally 
measured only sets a lower limit corresponding to $\overline\delta (x) =
\delta (x) = 0$.

We can summarise this letter by saying that, in principle, the discrepancy between
theory and experiment found by the NMC \cite{nmc} could come from 
flavour symmetry violation in the nucleon sea and charge symmetry 
breaking
in either the the nucleon sea or valence distributions. 
However, because of the enormous value for $\overline\delta$ needed to
fit the experiment with charge symmetry breaking alone, 
it is more likely that the NMC result implies
some strong flavour symmetry breaking in the nucleon sea with a small 
$\overline\delta$ admixture. 

Of course, there are
many successful calculations based on pion physics \cite{achar,wally,koepf,harald} that
give a clear indication of an excess of $\overline d$ over $\overline u$ 
in the nucleon. On the other hand, the interpretation
of the NMC experiment as a confirmation of broken sea symmetry, does not 
rule out the possibility that, at a particular $x$, charge symmetry
breaking between the neutron
and the proton may be at the same level as that of flavour symmetry breaking.  
Our analysis indicates that it is possible to have $\lambda_s (x)$ larger
than the value quoted by the NA51 group at the cost of some
charge symmetry breaking between the proton and the neutron at a particular $x$,
even if the integrated value of this correction is zero or nearly
zero. It is clearly an urgent matter to find experimental ways to separate
these two contributions. For now, 
the important feature to note is that the NA51 result should be seen as 
a lower limit for $\lambda_s (x)$ and not as an absolute value. 
\newline
\newline
This work was supported by the Australian Research 
Council and by CAPES (Brazil).

\addcontentsline{toc}{chapter}{\protect\numberline{}{References}}

\end{document}